\documentclass[prb,twocolumn,amsmath,showpacs]{revtex4}
\usepackage{graphicx}
\usepackage{amssymb}
\usepackage{citesort}
\usepackage{stmaryrd}

\begin{document}

%% ---------- TITLE ----------
\title{$\pi$-electron theory of transverse optical excitons in semiconducting
single-walled carbon nanotubes
}

\author{Zhendong Wang}

\author{Hongbo Zhao}
\altaffiliation[Current address: ]
 {Department of Physics, University of Hong Kong, Hong Kong, China}

\author{Sumit Mazumdar}
\affiliation{Department of Physics, University of Arizona, Tucson,
  Arizona 85721, USA}

\date{\today}

%
% 73.22.-f   Electronic structure of nanoscale materials: clusters,
% nanoparticles, nanotubes, and nanocrystals
%
% 78.67.Ch   Optical properties of Nanotubes
%
% 71.35.-y   Excitons and related phenomena
%%% ---------------------------------- 
%%          abstract
%%%----------------------------------

\begin{abstract} 
We present a quantitative theory of optical absorption
polarized transverse to the tube axes in semiconducting single-walled
carbon nanotubes. Within one-electron theory, transverse optical absorption
occurs at an energy that is exactly in the middle of the two lowest 
longitudinal absorption energies. For nonzero Coulomb interactions
between the $\pi$-electrons, transverse optical absorption
is to an exciton state that is strongly
blueshifted relative to the longitudinal excitons. 
Very similar behavior is observed in the $\pi$-conjugated polymer
poly-paraphenylenevinylene, where the optical absorption polarized
predominantly perpendicular to the polymer chain axis is blueshifted 
relative to the
absorptions polarized predominantly along the chain axis.
The binding energy of the transverse
exciton in the nanotubes is considerably smaller than those of the 
longitudinal excitons.
Electron-electron interactions also reduce the relative oscillator
strength of the transverse optical absorption.  Our theoretical results
are in excellent agreement with recent experimental measurements in four
chiral nanotubes.  
\end{abstract}

\maketitle

\section{Introduction}

It is now firmly established that optical absorptions polarized parallel
to the nanotube axes in semiconducting single-walled carbon nanotubes
(SWNTs) are to exciton states.
\cite{Ando97,Lin00,Kane,Marinopoulos,Spataru,Chang04,Zhao04,Perebeinos,Araujo07,
Saito07} Theoretical
work by different groups have established that in the subspace of excited
states that are coupled to the ground state by the longitudinal component
of the dipole operator, there occur a series of energy manifolds (labeled
by an index $n$ = 1, 2, ... etc.), with each energy manifold containing
optical and dark excitons, and a continuum band separated from the optical
exciton by a characteristic exciton binding energy. Nonlinear
spectroscopic measurements have been applied to determine the binding
energy of the $n=1$ longitudinal exciton (hereafter Ex1) in a number of SWNTs.
\cite{Wang05,Dukovic05,Maultzsch05,Zhao06}  The determination of the
binding energy of the $n=2$ exciton (hereafter Ex2) appears to be more difficult within
the existing experimental approaches.  Recent theoretical works have
suggested that electroabsorption may be a useful tool for this purpose.
\cite{Perebeinos07,Zhao07}

In contrast to the longitudinal absorption, the literature on optical
absorptions polarized transverse to the nanotube axes, theoretical
\cite{Ando97,Marinopoulos,Chang04,Zhao04,Uryu06} as well as experimental,
\cite{Miyauchi06,Lefebvre07} is rather limited. Early theoretical
investigations had largely emphasized the suppression of the
perpendicularly polarized absorption due to local field effects.
\cite{Ando97,Marinopoulos,Chang04} Later work based on $\pi$-electron
theory \cite{Zhao04} found nonzero oscillator strength for the transverse
absorption. More interestingly, while within the noninteracting nearest
neighbor (nn) tight-binding model the transverse absorption is expected at
the exact middle of the lowest two longitudinal absorptions, inclusion of
the Coulomb interactions between the $\pi$-electrons leads to a splitting
of the final states to which transverse absorption occurs: to a redshifted
forbidden state and a blueshifted allowed state.  \cite{Zhao04} Correlated
electron calculation for the (8,0) SWNT found the allowed transverse
absorption relatively far from Ex1 occurring at energy $E_{11}$ and very
close to Ex2 occurring at $E_{22}$. \cite{Zhao04} Significant blueshift of
the transverse absorption has also been found more recently within an
effective mass approximation theory. \cite{Uryu06}

Experimentally, excited states coupled to the ground state by the
transverse component of the dipole operator have recently been detected by
polarized photoluminescence excitation (PLE) spectroscopy.
\cite{Miyauchi06,Lefebvre07} Miyauchi {\it et al.} have determined the PL
spectra for four chiral SWNTs with diameters $d = 0.75 - 0.9$ nm.
\cite{Miyauchi06} In all cases, the allowed transverse optical absorption
is close to $E_{22}$, in agreement with theoretical prediction. \cite{Zhao04} Lefebvre
and Finnie have detected transverse absorptions close to $E_{22}$ in 25
SWNTs with even larger diameters, \cite{Lefebvre07} 
%These latter investigators 
and were also able to demonstrate the same ``family behavior'' in transverse
absorption energies that had been noted previously with longitudinal
absorptions. \cite{Bachilo02}

There is general agreement between the different research groups that the
extent of the blueshift of the transverse absorption is a direct measure
of the strength of the Coulomb electron-electron (e-e) interactions in
SWNTs. \cite{Zhao04,Uryu06,Miyauchi06,Lefebvre07} Indeed this blueshift is
much more easily determined (provided the states are observable
experimentally) than other correlation-induced energy splittings such as
exciton binding energies or the energy differences between the optical and
dark excitons within any longitudinal manifold. We have recently
demonstrated that the  Coulomb interaction parameters obtained by fitting
the 
%longitudinal and transverse absorptions 
absorption spectra in the $\pi$-conjugated
polymer poly-paraphenylenevinylene (PPV) \cite{Chandross97b} can reproduce
quantitatively the energies of Ex1, Ex2 and the corresponding two-photon
excitons in SWNTs with diameters $d \geq 0.75$ nm. \cite{Zhao06,Wang06}  Here we
report detailed calculations of the transverse absorption for the four
SWNTs that have been investigated by Miyauchi {\it et al.}, \cite{Miyauchi06} 
(6,5), (7,5), (7,6) and (8,4), within the same model
$\pi$-electron Hamiltonian with the same e-e and one-electron parameters.
Our goal is to establish the model Hamiltonian and the parameters firmly
for SWNTs. In addition, we point out the remarkable parallels between the
present discussions and earlier ones focusing on 
optical absorption polarized perpendicular
to the polymer chain axis in PPV  \cite{Chandross97a, Comoretto98, Comoretto00, 
Miller99,Kohler98, Rice94, Gartstein95, Chandross97b, Bursill02}.
Comparisons between PPV and SWNTs allow us to
make the effects of e-e interactions clearer.

\section{Theoretical model, parametrization and boundary condition}

Our calculations are within the  $\pi$-electron Pariser-Parr-Pople 
(PPP) model Hamiltonian, \cite{Pariser}
\begin{eqnarray}
H &=& H_{1e} + H_{ee} ,\nonumber \\
H_{1e} &=& -\sum_{i\neq j,\sigma}t_{ij}(c_{i\sigma}^{\dagger}c_{j\sigma} + c_{j\sigma}^{\dagger} c_{i\sigma}), \label{eqn-h}\\
H_{ee} &=& U \sum_i n_{i\uparrow}n_{i\downarrow} + \frac{1}{2} \sum_{i\neq j} V_{ij}(n_i-1)(n_j-1) . \nonumber
\end{eqnarray}
Here $H_{1e}$ and $H_{ee}$ consist of one-electron and many-electron interactions;
$c^\dagger_{i \sigma}$ creates a $\pi$-electron with spin
$\sigma$
on the $i$th carbon atom, 
$n_{i \sigma} = c^\dagger_{i \sigma}c_{i \sigma}$ is the number of
$\pi$-electrons with spin $\sigma$ on the atom $i$, and $n_i =
\sum_{\sigma} n_{i\sigma}$ is the total number of $\pi$-electrons on the
atom. 
The parameters $t_{ij}$ are the one-electron hopping integrals, $U$ is the
repulsion between two $\pi$-electrons occupying the same carbon atom, and
$V_{ij}$ the intersite Coulomb interactions.  In our previous work
\cite{Zhao04,Zhao06,Wang06} we had limited the electron hopping to nn
$t_1$ only; here we investigate the consequences of including next nearest
neighbor (nnn) hopping $t_2$, to determine the effects of broken
charge-conjugation symmetry (CCS).  Miyauchi {\it et al.}
\cite{Miyauchi06} have tentatively ascribed a splitting observed in the
transverse absoption bands of SWNTs to broken CCS; we examine this issue
theoretically.  Our choice of the Coulomb parameters, hereafter referred
to as screened Ohno parameters, \cite{Chandross97b} and the $t_1=2.0$ eV,
are the same as in the case of longitudinal excitations. \cite{Wang06} We
use the single configuration interaction (SCI) approximation, which
involves solving the Hamiltonian of Eq.~(1) within the subspace of one
electron-one hole excitations from the Hartree Fock ground state. The
justification for this approximation has been given in our previous work.
\cite{Zhao04,Zhao06,Wang06} Our calculations are for open boundary
condition, which enables precise determinations of transition dipole
moments. \cite{Zhao04,Zhao06,Wang06} The number of unit cells we retain
are 5 -- 6 for the (6,5), (7,5) and (7,6) SWNTs, and 22 for the (8,4)
SWNT, with 2100 -- 2500 carbon atoms in all cases.  Convergence in the
calculated energies as well as excellent agreement with experiment are
seen at these system sizes for the $n=1$ and 2 longitudinal optical
excitons. \cite{Wang06}

%% ----------------------------
%%          Fig. 1
%% ----------------------------
\begin{figure}
 \centering
 \includegraphics[clip,width=2.45in]{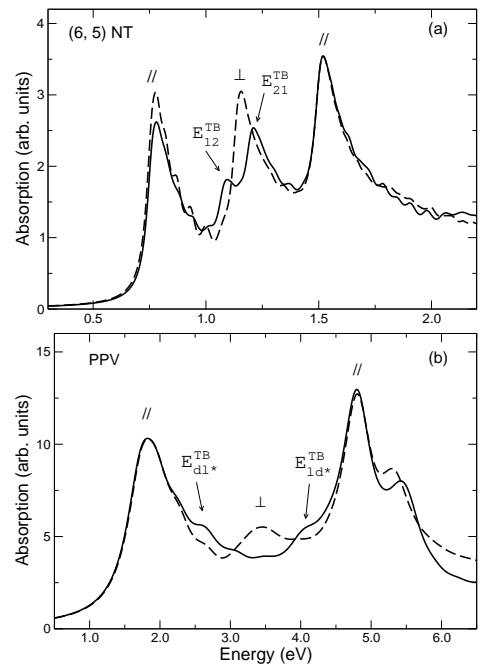}
 \caption{
  Calculated optical absorption spectra of (a) a (6,5) SWNT, and (b) PPV within
the tight binding model for $t_2=0$ (dashed line) and $t_2=0.6$ eV (solid line). The
longitudinal  ($\sslash$) and transverse ($\perp$) components of the optical absorption 
are indicated in each case.
} 
\end{figure}

\section{Results and discussion}

\subsection{One-electron tight-binding limit}

In view of what follows we begin with a discussion of the $U=V_{ij}=0$
tight binding (TB) limit of Eq.~(1) for both SWNTs and PPV. Transverse
absorption in SWNTs is due to the optically induced transitions from the
highest valence subband $v_1$ to the second lowest conduction subband
$c_2$, and from the second highest valence band $v_2$ to the lowest
conduction band $c_1$ (see Fig.~2 in Ref.~\onlinecite{Zhao04}).  We denote
the one-electron excitations as $\psi_{v1 \to c2}$ and $\psi_{v2 \to c1}$,
respectively, and the corresponding excitation energies as
$E^\text{TB}_{12}$ and $E^\text{TB}_{21}$.  The two transitions are
degenerate for $t_2=0$, and occur exactly at the center of the two
longitudinal transitions $E^\text{TB}_{11}$ and $E^\text{TB}_{22}$.  This
is shown in Fig.~1(a) for the (6,5) SWNT, for $t_1=2.0$ eV. The nn only TB
band structure of PPV is similar (see Fig.~1 in
Ref.~\onlinecite{Chandross94}), although the nomenclature is different.
The equivalents of $v_2$ and $c_2$ are localized flat bands which are
labeled $l$ and $l^*$, respectively, while the delocalized equivalents of
$v_1$ and $c_1$ are labeled $d$ and $d^*$, respectively. Once again the
degenerate transverse transitions $d \to l^*$ and $l \to d^*$ occur
exactly at the center of the longitudinal $d \to d^*$ and $l \to l^*$
transitions, as is shown in Fig.~1(b) for $t_1=2.4$ eV, which is
appropriate for planar $\pi$-conjugated systems.  The oscillator strength
of the central peak, relative to those for the longitudinal transitions,
is much larger in the SWNT than in PPV.  This is a reflection of the
larger electron-hole separation that is possible in the transverse
direction in a SWNT with $d \sim 1$ nm, as compared to PPV. 

The degeneracy between $E^\text{TB}_{12}$ and $E^\text{TB}_{21}$
($E^\text{TB}_{dl^*}$ and $E^\text{TB}_{ld^*}$ in PPV) is lost if CCS is
broken by including $t_2$. Fig.~1 shows the effect of $t_2=0.6$ eV on the
absorption spectra of the (6,5) SWNT and PPV; we will argue below that
this is the largest possible nnn hopping between $\pi$-orbitals. The
splitting between the transverse transitions is much smaller in the (6,5)
SWNT than in PPV. We have found this to be true for all four SWNTs we have
studied.  We do not show results of including a third-neighbor hopping,
since this does not contribute any further to the splitting of the
transverse states.

%% ----------------------------
%%          Fig. 2
%% ----------------------------
\begin{figure}
 \centering
 \includegraphics[clip,width=3.375in]{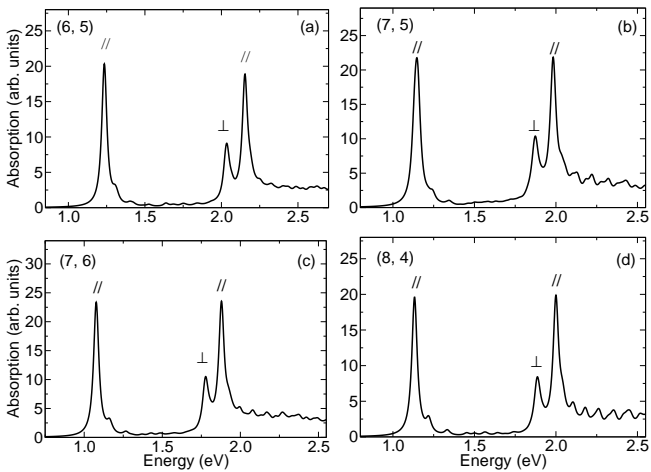}
 \caption{Calculated optical absorption spectra within the PPP model,
with $t_1=2.0$ eV, $t_2=0$, for four chiral SWNTs. The longitudinal
and transverse components of the absorption are indicated in each case.
}
\end{figure}

\subsection{Nonzero e-e interaction}

We now discuss the full PPP Hamiltonian of Eq.~(1).  The matrix element
$\langle \psi_{v1 \to c2}|H_{ee}|\psi_{c1 \to v2} \rangle$ is nonzero, and
as a consequence eigenstates of the Hamiltonian are now odd and even
superpositions of these basis functions, $\Psi_O = \psi_{v1 \to
c2}-\psi_{c1 \to v2}$ and $\Psi_E = \psi_{v1 \to c2}+\psi_{c1 \to v2}$.
Repulsive $H_{ee}$ ensures that the optically forbidden $\Psi_O$ is
redshifted while the optically allowed $\Psi_E$ is blueshifted.  
%%SM8/30 - new sentence
The mechanism of the above energy splitting into bright and dark states
is identical to that behind the splitting of the
longitudinal excitons into bright and
dark excitons \cite{Zhao04}, with the only difference that the magnitude of the
splitting in the present case is much larger (see below). 
This strong first order configuration interaction effect occurs over and
above the overall blueshift of the entire absorption spectrum due to e-e
interactions; as a consequence the allowed transverse exciton $\Psi_E$ is
blueshifted also with respect to the longitudinal excitons. 
In Fig.~2 we have shown our calculated optical absorption spectra within
Eq.~(1) for $t_1=2.0$ eV, $t_2=0$, for all four SWNTs investigated by
Miyauchi {\it et al.} \cite{Miyauchi06}  In all cases the optically
allowed transverse exciton is seen to occur very close to the $n=2$
exciton, as observed experimentally. \cite{Miyauchi06} The relative
oscillator strength of the transverse exciton is now considerably weaker
than Ex1 and Ex2. 

The mechanism of the splitting of the degenerate
transverse excitations is identical to that in PPV, where also the
optically allowed $d \to l^* + l \to d^*$ transverse excitation occurs
closer to the $l \to l^*$ exciton than to the $d \to d^*$ exciton.
\cite{Chandross97b,Chandross97a,Rice94,Gartstein95,Bursill02} We emphasize
that the coupling between $\psi_{v1 \to c2}$  and $\psi_{c1 \to v2}$ 
is independent of the
boundary condition (periodic versus open) along the longitudinal direction: the calculations
for PPV in references \onlinecite{Rice94} and \onlinecite{Gartstein95}, for example, 
employ periodic
boundary condition. 
%%SM8/30 - new
In the present case, we have repeated our calculations of all energies and
wavefunctions, but not transition dipole couplings (which is difficult to
define with periodic boundary condition within tight-binding models
\cite{Gonokami94}), for for all four SWNTs also with periodic boundary
condition. In every case we have confirmed the splitting of the transverse
wavefunctions into $\Psi_O$ and $\Psi_E$ from wavefunction analysis. We
have confirmed that the energy differences between the odd and even
superpositions the same with the two boundary conditions, for the number
of unit cells used in the calculation.  Our parametrization of the
$V_{ij}$ involves a dielectric constant \cite{Zhao04,Wang06}. The energy
splitting between the odd and even superpositions will occur for any
finite dielectric constant, and only the magnitude of the splitting
depends on the value of the dielectric constant.

We have previously shown that the screened Ohno Coulomb parameters along
with $t_1=2.0$ eV and $t_2=0$ reproduce the experimental $E_{11}$ and
$E_{22}$ of the four SWNTs of interest quantitatively.  The difference
between the experimental and calculated $E_{11}$ is 0.02 -- 0.06 eV, while
the same for $E_{22}$ is about twice these values (see Table II in
Ref~\onlinecite{Wang06}).  In Fig.~3 we make quantitative comparisons
between the experimental transverse optical absorption spectra of Miyauchi
{\it et al.} \cite{Miyauchi06} with those calculated within Eq.~(1) for
both $t_2=0$ and $t_2=0.6$ eV.  As seen in Fig.~1, $E_{11}$ and $E_{22}$
in SWNTs are unaffected by $t_2$ even for $H_{ee}=0$, so that our fits to
these \cite{Wang06} continue to be valid. The low frequency regions of the
experimental spectra are dominated by background noise and are therefore
ignored in our discussions below.

%% ----------------------------
%%          Fig. 3
%% ----------------------------
\begin{figure}
 \centering
 \includegraphics[clip,width=3.375in]{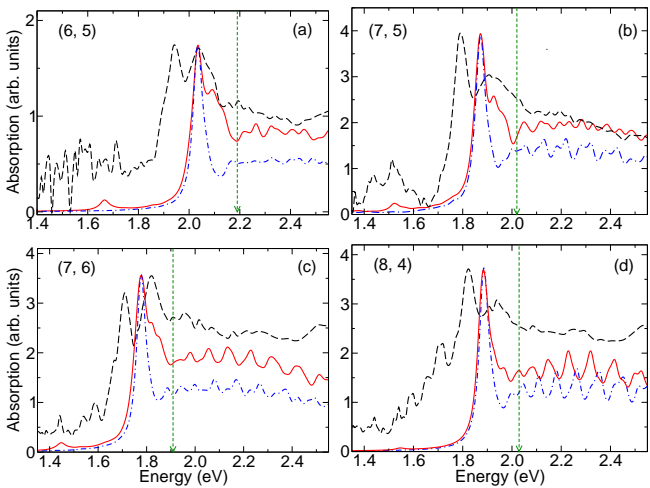}
 \caption{
   (Color online) 
    Comparison of experimental\cite{Miyauchi06} (dashed curves) and calculated 
    transverse components of optical absorptions in four SWNTs for $t_2=0$ (dot-dashed curves) 
    and $t_2=0.6$ eV (solid curves). The vertical lines correspond to the Hartree--Fock
threshold.
}
\end{figure}

We compare experimental and calculated transverse optical absorptions of
the SWNTs in Fig.~3. The peak heights of the calculated absorption
spectra in Fig.~3 have been adjusted to match those of the experimental
spectra. The experimental absorption spectra show two peaks with nearly
the same separation in all cases, $\sim$ 0.1 eV.  Independent of which of
these two peaks correspond to the true electronic energy of $\Psi_E$, it
is clear that the error in our calculation is again small, $\leq 0.1$ eV.
The experimental transverse absorption bands are much broader than the
longitudinal absorption bands in all cases, a feature that is
qualitatively reproduced in our calculated spectra. This is a consequence
of the much smaller binding energy of the transverse exciton, which leads
to incomplete transfer of oscillator strength from the continuum band to
the exciton.  Binding energies of longitudinal excitons are larger and
there occur greater transfer of oscillator strength from the corresponding
continuaa to the excitons. \cite{Miyauchi06,OConnell02} The broad
absorption beyond the excitons in Fig.~3 are therefore to continuum band
states. We have indicated in Fig.~3 the Hartree-Fock thresholds for the
transverse states, which is the lower edge of the continuum band within
the SCI approximation. The binding energies of the transverse excitons,
taken as the difference between the Hartree-Fock threshold and the exciton
energy, are $\sim$ 0.15 eV for all four SWNTs, which is about one-third of
that of the $n=1$ longitudinal exciton.  Miyauchi {\it et al.}, from a
different perpsective, have also arrived at the conclusion that the
binding energy of the transverse exciton is small. \cite{Miyauchi06}
Similar conclusion was reached in PPV from photoconductivity
studies. \cite{Kohler98}

We now discuss the energy splitting of $\sim$ 0.1 eV between the peaks in
the experimental absorption spectra.  Miyauchi {\it et al.} ascribe this
to broken CCS, viz., nondegenerate $E^\text{TB}_{12}$ and
$E^\text{TB}_{21}$ even within $H_{1e}$. Our calculated absorption spectra
in Fig.~3 for $t_2$ as large as 0.6 eV, nearly one-third of $t_1$,
however, fail to reproduce this splitting. This is to be anticipated from
the $H_{ee}=0$ absorption spectrum of Fig.~1(a), since any splitting due
to broken CCS, a one-electron effect, can only be smaller for $H_{ee} \neq
0$. At the same time, $t_2=0.6$ eV should be considered as the upper limit
for the nnn electron hopping based on experiments in PPV as we explain
below.
 
%%%%%%%%%%%%%%%%%%%%%%%%%%%%%%%%%%%%%% %SM5/10 - rewriting
%%%%%%%%%%%%%%%%%%%%%%%%%%%%%%%%%%%%%%
The experimental absorption spectra of PPV-derivatives resemble {\it
qualitatively} the tight-binding absorption spectrum of Fig.~1(b) for
$t_2\neq0$, with, however, an overall blueshift due to e-e interactions. The
spectra contain strong absorption bands at $\sim$ 2.2 -- 2.4 eV and 6 eV,
and weaker features at 3.7 eV and 4.7 eV, respectively. The lowest and
highest absorption bands are polarized predominantly along the polymer
chain axis, while the 4.7 eV band is polarized perpendicular to the chain
axis \cite{Chandross97a,Comoretto98,Comoretto00,Miller99}. It is then
tempting, based on Fig.~1(b), to ascribe the origin of the 3.7 eV band to
broken CCS, in which case it ought to have the same polarization as the
4.7 eV absorption band.  Repeated experiments have, however, found that
the absorption band at 3.7 eV to be polarized predominantly along the
polymer chain axis.  \cite{Chandross97a,Comoretto98,Miller99,Comoretto00}
Theoretical calculations within the PPP model with $t_2=0$ have reproduced
the longitudinal polarization of the 3.7 eV band, which is ascribed to the
second lowest longitudinal exciton in PPV derivatives.
\cite{Chandross97b,Chandross97a,Bursill02} We have confirmed that
inclusion of $t_2=0.6$ eV within the same PPP model calculation renders
the polarization of the the 3.7 eV absorption band perpendicular to the
polymer chain direction (not shown), in contradiction to experiments. The
nnn hopping in PPV is therefore certainly smaller than 0.6 eV.  Curvature
of SWNTs implies an even smaller value in SWNTs. \cite{Wang06} We are
consequently unable to give a satisfactory explanation of the splitting in
the transverse absorption in SWNTs. We have not found any higher energy
transverse exciton whose oscillator strength can explain the observed
absorption spectra. It is conceivable that the second peak in the
experimental absorption spectra in Fig.~3 correspond to the threshold of
the transverse continuum band (see in particular the spectra in Figs. 3(b)
and (d)). Further experimental work is therefore warranted. It is also
possible that the energy splitting is due to higher order correlation
effects neglected in SCI, or due to intertube interactions. We are
currently investigating the latter possibilities.

\section{Conclusion}

In conclusion, eigenstates coupled to the ground state by the transverse
component of the dipole operator are degenerate within one-electron theory
in both SWNTs and PPV. $H_{ee} \neq 0$ splits this degeneracy, and the
optically allowed higher energy state appears close to the higher energy
longitudinal excitons in SWNTs. The binding energies of the transverse
excitons are about one-third of those of the longitudinal excitons. The
quantitative aspect of our calculations is worthy of note. 
%Taken together with our earlier calculations \cite{Wang06}, 
We have now demonstrated that the same model Hamiltonian with the same
one-electron hopping and Coulomb interactions can reproduce the
experimental energies and absorption spectra of longitudinal and
transverse optical excitations in SWNTs with diameters greater than 0.75
nm with considerable precision (errors $\leq$ 0.05 -- 0.1 eV).  In those
cases where the experimental binding energies of Ex1 are known, the
calculated quantities are uniformly very close. \cite{Wang06} It has been
suggested that the true single tube binding energies are considerably
larger than the 0.3 -- 0.4 eV that are found experimentally
\cite{Wang05,Dukovic05,Maultzsch05,Zhao06} for SWNTs with diameters
$\sim$ 0.75 -- 1 nm, and the experimental quantities reflect strong
screening of e-e interactions by the environment. The close agreements between our
theoretical single tube calculations and experiments suggest, however,
that any such environment effect on the exciton binding energy is small. 

\begin{acknowledgments}
We thank the authors of Ref. \onlinecite{Miyauchi06} for sending us the
data used to construct the experimental spectra of Fig.~3. This
work was supported by NSF-DMR-0705163.
\end{acknowledgments}

\end{document}